\documentclass[superscriptaddress, prd, aps,amsmath,amssymb,showpacs,showkeys, onecolumn]{revtex4-2}
\usepackage[dvips]{graphicx,color}
\usepackage{subfig}
\usepackage{mathrsfs}
\usepackage{times}
\usepackage{xcolor}
\usepackage[%
  colorlinks=true,
  urlcolor=blue,
  linkcolor=red,
  citecolor=blue
]{hyperref}
\usepackage{orcidlink}

\begin{document}
\title{Quantum-Corrected Thermodynamics and Phase Structure of AdS Euler–Heisenberg Black Hole}

\author{Siddhartha Sankar Borah \orcidlink{0000-0003-0126-8941}}
\email[Email: ]{siddharthasborah95@gmail.com} 
\affiliation{%
 Department of Physics, Dibrugarh University, Dibrugarh \\
 Assam, India, 786004}%

\author{Dhruba Jyoti Gogoi \orcidlink{0000-0002-4776-8506}}%
 \email[Email: ]{moloydhruba@yahoo.in}
\affiliation{Department of Physics, Madhabdev University, Narayanpur, Lakhimpur 784164, Assam, India}
\affiliation{Research Center of Astrophysics and Cosmology, Khazar University, Baku, AZ1096, 41 Mehseti Street, Azerbaijan}
 
\author{Kalyan Bhuyan\orcidlink{0000-0002-8896-7691}}%
 \email[Email: ]{kalyanbhuyan@dibru.ac.in}
\affiliation{%
 Department of Physics, Dibrugarh University, Dibrugarh \\
 Assam, India, 786004}%
 \affiliation{Theoretical Physics Division, Centre for Atmospheric Studies, Dibrugarh University, Dibrugarh, Assam, India 786004}

\begin{abstract}
We investigate the thermodynamic behaviour of an AdS black hole arising from nonlinear electrodynamics-corrected gravity, incorporating quantum effects through thermal fluctuations. From the Einstein--Euler--Heisenberg framework, we consider the modified black hole solution and analyse its thermodynamic properties in the extended phase space. Logarithmic and inverse-area corrections to the entropy are obtained, leading to modified expressions for enthalpy, internal energy, Helmholtz free energy and Gibbs free energy. The corrected specific heat exhibits multiple divergences and sign changes, signalling genuine second-order phase transitions and revealing a quantum-stabilized microscopic phase followed by universal macroscopic instability. Our results demonstrate that thermal fluctuations qualitatively restructure the thermodynamic phase space and highlight the dominant role of quantum corrections in governing the black hole stability.
\end{abstract}

\keywords{Einstein--Euler--Heisenberg Gravity; Corrected Thermodynamics; Black holes; Entropy}

\maketitle
\section{Introduction}

In recent years, studies on black holes have gained significant importance not only from a gravitational perspective but also due to their intriguing thermodynamic behavior. Black holes and their thermodynamic framework are a combination of the theories of classical thermodynamics, general relativity, quantum mechanics, and statistical mechanics, emerging as a fundamental area of research in theoretical physics. This field aims to apply the thermodynamic principles to black holes to have a more vivid picture of the current understanding of our universe. J.D. Bekenstein in the year 1973 first tried to correlate the concept of entropy with black holes, where he argued that black holes possess entropy proportional to the area of their event horizon, thereby regarding black holes as thermodynamic objects\cite{PhysRevD.7.2333}. But any matter falling into a black hole is lost forever, leading to a loss of entropy and potentially violating the second law of thermodynamics. To resolve this dogma, Bekenstein introduced the concept that black holes must themselves have entropy.

Based on Bekenstein’s proposition, S. Hawking in 1974 demonstrated that black holes emit thermal radiation \cite{PhysRevLett.105.203901} now known as Hawking radiation with temperature proportional to their surface gravity\cite{Bekenstein:1972tm}. This helped establish that black holes are not entirely isolated objects but also indicates they lose mass and energy over time, thereby challenging the conventional understanding that nothing can escape a black hole. Hawking also proved that the surface area of a black hole’s event horizon cannot decrease\cite{PhysRevD.7.2333}, thereby further strengthening the thermodynamic comparison. These developments eventually led to the formulation of the four laws of black hole thermodynamics, which closely mirror the classical laws of thermodynamics\cite{Bardeen:1973gs}.

A key feature characterizing black hole thermodynamics is the Bekenstein-Hawking area law, which states that the entropy S of a black hole is related to its area of event horizon as S=A/4\cite{PhysRevD.7.2333, Bardeen:1973gs}. This law is primarily applicable to large black holes in equilibrium, but for smaller black holes, whose equilibrium gets affected by Hawking radiation, it requires correction to the area law, thereby necessitating correction to the entropy. As a result, a logarithmic term needs to be introduced as a first-order correction to the entropy\cite{2023_Mandal}.These corrections, derived from various statistical and CFT methods\cite{article}, have dynamic effects on the thermodynamic behavior and black hole stability. These corrections to the black hole entropy have led to modification in various thermodynamic parameters such as temperature, specific heat, enthalpy, and free energy\cite{doi:10.1142/S0218271823501109,Upadhyay:2018gee,10.1093/ptep/pty145,Wald:1999vt,Upadhyay:2018vfu,UPADHYAY2017130,doi:10.1142/S0217732321502126, Bora:2025zja, Gogoi:2025rcn}. Recent studies have investigated the effects of these corrections on different black hole solutions, including rotating and charged black holes in anti-de Sitter (AdS) spaces, dilatonic black holes, and Schwarzschild-Beltrami-de Sitter black holes\cite{Upadhyay:2018gee,cmp/1103922135,Pourhassan:2017rie,cmp/1103922135}. It has been observed that first-order corrections can destabilize small black holes, but second-order corrections often enhance their stability\cite{256e337715504f60b40bfb83a385bbbb,
Gogoi:2024ypn}. Moreover, external fields and dark energy components like quintessence and cosmological constants have been found to significantly influence the thermodynamic properties of black holes, leading to phase transitions akin to those in classical thermodynamical systems. 

Applying inputs from the holography principle, in particular AdS/CFT correspondence, this field of study has gained more significance. Pourhassan et al \cite{Pourhassan:2016atz,Pourhassan:2016zzc}. studied the behavior of rotating Kerr-AdS black holes, black Saturns, and other such complex systems within this framework, providing valuable insights into our current understanding of black hole microstructures and phase transitional behavior. In light of such developments, Chamblin et al. \cite{Chamblin:1999hg} investigated the parameters such as holography, thermodynamics, and quantum fluctuations. According to Hawking and Page’s contribution to black hole studies, when Hawking temperature reaches a critical point, AdS Schwarzschild’s black hole undergoes a phase shift\cite{cmp/1103922135}. As a result of this transition, P-V criticality of the black holes has also become a topic of interest, along with first-order correction to dumb holes (black hole counterparts)\cite{Pourhassan:2016xsy}. It is worth mentioning that logarithmic thermodynamic adjustments do not affect the stability of modified  Hayward’s black hole\cite{Pourhassan:2016qoz}.
Certain black holes, such as the Bardeen black hole and the modified Hayward black hole, involve nonlinear electrodynamics or coupling with dark energy components such as quintessence, showing that quantum corrections may or may not affect stability, depending on the black hole structure and external conditions. Recent studies have also shown that while small black holes are sensitive to thermal fluctuations leading to a decrease in entropy and violation of the second law, larger black holes remain largely unaffected\cite{256e337715504f60b40bfb83a385bbbb}. In light of these developments, C.H Nam studied the higher-order entropy corrections for a non-linear magnetic-charged AdS black hole surrounded by a quintessence field, where he showed that at a certain critical temperature, a black hole may undergo a thermal phase transition from a larger unstable black hole to a stable smaller black hole\cite{Nam:2018uvc}.
In Ref.\cite{Singh:2025ueu}, the authors have studied how charged AdS black holes are affected by the presence of a string cloud, which acts as an additional gravitational field and modifies the space-time geometry and the behavior of black hole thermodynamics at the boundary CFT. The authors reported that at this stage, heat capacity at constant charge becomes infinite at a certain point, indicating a second-order phase transition similar to what happens at the critical points in fluids \cite{Kubiznak:2012wp,Shen:2005nu,Sahay:2010tx}. Here, pressure instead of volume changes during the phase transition, suggesting a special behavior in boundary CFT. The authors have also calculated the thermodynamic scalar curvature (R) of the charged AdS black hole, which helps to identify the type of interaction between the black hole micromolecules and bridge the gap in understanding quantum gravity. In Ref.\cite{Czinner:2025koi}, the authors have discussed the long-standing inadequacy of systems with long-range interactions and strong gravitational fields, where the standard Boltzmann-Gibbs (BG) entropy based on the Hawking-Bekenstein area law leads to non-additivity and thermodynamic instability, with special reference to Schwarzschild black holes. To address this issue, generalized entropies such as Tsallis entropy \cite{Tsallis:2009zex} and Rényi entropy \cite{Rényi1959} have been proposed. The latter is more compatible, as it offers the advantage of being additive and more relatable with the zeroth law of thermodynamics, while still occupying the non-extensive effects through the parameter \( \lambda \) as seen in the Renyi entropy formula \cite{36a855f1-e690-375a-abef-ca42ce206b16} $S_R = -\frac{1}{\lambda} \ln \sum p_i^{1 - \lambda}$  which is related to the Tsallis entropy as $S_R = \frac{1}{\lambda} \ln [1 - \lambda S_T]$. Based upon the insights of applying Renyi entropy to black hole thermodynamics, the authors in this paper have constructed a black hole solution where the Hawking temperature exactly matches the thermodynamic temperature derived from Renyi formalism. This is done by introducing an anisotropic fluid surrounding the black hole and described by a Kiselev-type metric \cite{unknown} where the fluid properties are unique to the black hole mass and Renyi parameter. The solutions smoothly converge to standard relativity as the result holds when $\lambda \to 0$. The study also explains that the Schwarzschild scenario remains thermodynamically unstable; the third law of thermodynamics imposes an upper bound on the black hole mass, effectively acting as a cosmic censor by ensuring the preservation of the event horizon. In Ref.\cite{Joy:2025hnk}, the authors have discussed the thermodynamic properties—such as temperature, entropy, and heat capacity—of an advanced  black hole model known as charged Reissner-Nordström Black Bounce Black Holes, a concept first introduced by Simpson and Visser. These models are devoid of any singularities and feature a minimum surface at the center, making them mathematically feasible and serving as a potential candidate for exotic phenomena like "traversible wormholes." This model incorporates both electric charge and an anisotropic fluid, combining standard Maxwell electromagnetism with these alternative geometries. Black bounce black hole models are seen as black hole mimickers since they can reproduce observable effects like shadows and gravitational lensing similar to that of classical black holes \cite{Zhao:2023vxq} while still avoiding singularities. It is seen that the quantum effect dominates at smaller scales, while thermal properties become more influential as the black hole grows in size.

Classically, the relation for calculating the entropy of black holes is  given by the Hawking-Bekenstein area law $S=A/4$ establishing entropy as proportional to horizon area, but quantum gravity predicts important corrections. Among these, logarithmic and exponential corrections become significant. Logarithmic corrections arise universally from one-loop quantum fluctuations around the black hole background and have been derived in various approaches, including string theory, loop quantum gravity, and quantum field theory on curved spacetime\cite{Ahluwalia_2017,Fernando:2006gh}. Non-perturbative effects such as instantons, topological configurations, or strong-field matter interactions can introduce exponential corrections \cite{Banerjee:2008gc,Solodukhin:2011gn}.So the corrected entropy formula becomes \( S = S_{\mathrm{BH}} + \alpha \log S_{\mathrm{BH}} + \gamma e^{-\delta S_{\mathrm{BH}}} \). Here, the terms $\gamma$,$\delta$ are non-perturbative and are physically related to brane effects, non-local entanglement, instantons and do not appear in the finite order expansion of $1/S_H$. 
The exponential corrections to black hole entropy arise primarily from non-perturbative effects within the Euclidean path integral approach to quantum gravity. This exponential term reflects the contributions from the subdominant, non-trivial saddle points with finite Euclidean action (referred to as instantons), which capture the quantum tunnelling 
processes between distinct vacua or topological sectors of the gravitational field, thereby introducing terms of the form
$\gamma e^{-\delta S_{\mathrm{BH}}}$ in the entropy expansion. 
Additionally, in theories with specific matter couplings, such as nonlinear electrodynamics (NED) \cite{article445} described by the Euler–Heisenberg Lagrangian, the effective action admits non-perturbative solutions that further contribute to these exponentially suppressed corrections. Collectively, such terms encode rich quantum gravitational phenomena beyond perturbative expansions,  which necessitates a comprehensive treatment accounting for both local quantum fluctuations and global topologically nontrivial configurations in the black hole path integral. The Euler-Heisenberg nonlinear electrodynamics (EEH) \cite{Heisenberg:1936nmg,Weisskopf:1936hya,Schwinger:1951nm} is an effective field theory that describes the quantum corrections to classical electrodynamics arising from vacuum polarization effects in strong electromagnetic fields, leading to nonlinear interactions among electromagnetic fields due to virtual electron-positron pairs. Similarly, the Dirac-Born-Infeld (DBI) theory \cite{Gibbons:2001sx,Born:1934gh} also modifies Maxwell’s equations by limiting field strengths to avoid singularities. In \cite{Gursel:2025wan}, the authors have studied the exponential correction to black hole entropy in the context of Euler-Heisenberg nonlinear electrodynamics (NED) by considering the black hole as a thermodynamic system of many micro-particles, where the number of distinct permutations depends on the repetition frequency of each microstate configuration. The total number of microparticles and the total energy are calculated by summing over microstates with occupation numbers and associated energies. By maximizing the entropy (defined as the logarithm of the number of microstates), the most probable distribution of repetitions as an exponential function of a variation parameter is obtained under constraints on the total number of particles and energy. Enforcing these constraints and applying the large-N approximation, an expression for the exponentially corrected black hole entropy is obtained as the sum of the original entropy and a correction term reflecting the underlying quantum statistical nature of the microstates on the thermodynamic properties of black holes.

{
While the study in \cite{Gursel:2025wan} provides valuable insights into the quantum-corrected thermodynamics of the asymptotically flat Euler-Heisenberg black hole, it does not incorporate a cosmological constant. In contrast, our present work distinguishes itself by focusing on the anti-de Sitter (AdS) Euler-Heisenberg black hole, where a negative cosmological constant ($\Lambda < 0$) is introduced. This critical addition fundamentally alters the system, introducing a rich thermodynamic behavior that includes an extended phase structure (where $\Lambda$ is treated as thermodynamic pressure) and modified stability conditions. Furthermore, our study systematically investigates the interplay between the AdS background and thermal fluctuations, specifically exploring the effects of logarithmic and inverse-area entropy corrections on the extended phase transitions and critical behavior, phenomena which were not addressed in the asymptotically flat framework of \cite{Gursel:2025wan}.
}

From the above discussion, we aim to derive corrected expressions for key thermodynamic parameters such as enthalpy, Helmholtz free energy, thermodynamic volume, specific heat, internal energy, and Gibbs free energy, incorporating the effects of thermal fluctuations around equilibrium. This work will help contribute to a deeper understanding of how quantum and thermal fluctuation modifies the thermodynamic behavior of black holes within modified theories of gravity, offering insights into their stability and phase structure.

Throughout the whole paper, we have used $G=c=1$.

\section{AdS Black Hole Solutions in Nonlinear Electrodynamics-Corrected Gravity}
\label{sec2}

The intersection of quantum field theory and gravitation suggests that the standard Maxwell Lagrangian requires modifications in strong-field regimes. By incorporating one-loop quantum electrodynamics (QED) corrections via the Euler--Heisenberg effective Lagrangian, an effective nonlinear electrodynamics (NED) framework can be established. Following the approach by Ruffini et al. \cite{Ruffini:2013hia}, the effective Lagrangian is defined as:
\begin{equation}
 \mathscr{L}_{\text{eff}} = \mathscr{L}_M + \Delta \mathscr{L}_{\text{eff}}, 
\label{hey1}
\end{equation}
where $\mathscr{L}_M$ denotes the classical Maxwell term and $\Delta \mathscr{L}_{\text{eff}}$ represents the QED-induced correction. The real part of this perturbative correction is given by:
\begin{equation}
\left(\Delta\mathcal{L}^{\cos}_{\text{eff}}\right)_\mathcal{P} = \frac{1}{8\pi^2} \sum_{n,m=-\infty}^{\infty} \frac{1}{\tau_m^2 + \tau_n^2} \left[ \bar{\delta}_{m0} J(i\tau_m m_e^2) - \bar{\delta}_{n0} J(\tau_n m_e^2) \right].
\end{equation}
In this expression, the parameters are defined as $\tau_n = n\pi / (e\epsilon)$ and $\tau_m = m\pi / (e\epsilon)$, while $\bar{\delta}_{ij} = 1 - \delta_{ij}$ serves as the complementary Kronecker delta. The integral function $J(z)$ is expressed in terms of the exponential integral $\mathrm{Ei}(z)$ as:
\begin{equation}
J(z) \equiv \mathcal{P} \int_0^{\infty} ds \, \frac{s e^{-s}}{s^2 - z^2} = -\frac{1}{2} \left[ e^{-z} \mathrm{Ei}(z) + e^z \mathrm{Ei}(-z) \right].
\end{equation}
Here, the principal value $\mathcal{P}$ is used for the exponential integral $\mathrm{Ei}(z) \equiv \mathcal{P} \int_{-\infty}^{z} \frac{e^t}{t} \, dt$. 

In the weak-field limit ($\epsilon/E_c \ll 1$, $\xi/E_c \ll 1$), we utilize a perturbative expansion of the correction term:
\begin{equation}
\left(\Delta \mathcal{L}^{\cos}_{\text{eff}}\right)_\mathcal{P} = \frac{2\alpha^2}{45 m_e^4} \left(4 X^2 + 7 Y^2\right) + \mathcal{O}(\alpha^3).
\end{equation}
Consequently, the Euler--Heisenberg Lagrangian $\mathcal{L}_{\text{EH}} = X + (\Delta \mathcal{L}^{\cos}_{\text{eff}})_\mathcal{P}$ is incorporated into the Einstein--Euler--Heisenberg action:
\begin{equation}
\mathcal{I}_{EEH} = \int d^4x \sqrt{-g} \left[ -\frac{R}{16\pi G} + \mathcal{L}_{\text{EH}} \right]. 
\label{izm1}
\end{equation}

The resulting field equations are found by solving the Einstein equations $G^{\mu\nu} = 8\pi G T^{\mu\nu}$, where the energy-momentum tensor $T^{\mu\nu}$ is derived from the variation of the action $\mathcal{I}_{EEH}$ with respect to the metric. For a slowly varying electromagnetic field, $T^{\mu\nu}$ takes the form:
\begin{equation}
T^{\mu\nu} = T_M^{\mu\nu}(1 + \mathcal{A}_X) + g^{\mu\nu} \left[ \mathcal{A}_X X + \mathcal{A}_Y Y - \left(\mathcal{L}^{\cos}_{\text{eff}}\right)_\mathcal{P} \right], 
\label{hey6}
\end{equation}
where $\mathcal{A}_X$ and $\mathcal{A}_Y$ are derivatives of the Lagrangian with respect to the invariants $X$ and $Y$. In a purely electrostatic scenario ($B = 0$), the invariants simplify to $Y = 0$ and $X = E^2/2$, reducing the correction to:
\begin{equation}
\left(\Delta\mathcal{L}^{\cos}_{\text{eff}}\right)_\mathcal{P} = \frac{8\alpha^2}{45 m_e^4} X^2 + \mathcal{O}(\alpha^3). 
\label{hey3}
\end{equation}

By assuming a static, spherically symmetric metric of the form $ds^2 = f(r)dt^2 - f(r)^{-1}dr^2 - r^2 d\Omega^2$, we solve for the metric function $f(r)$. Utilizing the Einstein tensor component $G^{00}$ and the mass function $m(r)$, we determine the gravitational impact of the NED corrections. Following the normalization of the charge $Q$ as suggested in \cite{Magos:2020ykt, Luo:2022gdz}, the metric function for the Euler--Heisenberg black hole is given by \cite{Ruffini:2013hia}:
\begin{equation}
f(r) = 1 - \frac{2M}{r} + \frac{Q^2}{r^2} \left(1 - \frac{\mu Q^2}{20 r^4} \right). 
\label{nmetric}
\end{equation}
To investigate the properties of such objects in an expanding or anti-de Sitter background, we incorporate the cosmological constant $\Lambda$. The finalized metric function for a NED-corrected AdS black hole is expressed as \cite{Magos:2020ykt}:
\begin{equation}
f(r) = 1 - \frac{2M}{r} + \frac{Q^2}{r^2} \left(1 - \frac{\mu Q^2}{20 r^4} \right) - \frac{\Lambda r^2}{3}. 
\label{metricfunction}
\end{equation}
This solution serves as the geometric background for our subsequent analysis of the dynamic and thermodynamic characteristics of these modified black hole spacetimes.

\section{Basic Thermodynamic Properties}
\label{section3}

In the extended phase space formalism of black hole thermodynamics, the cosmological constant $\Lambda$ is treated as a thermodynamic pressure, thereby promoting the black hole mass $M$ to the role of enthalpy rather than internal energy. This formulation allows a direct analogy between asymptotically AdS black holes and ordinary thermodynamic systems. The thermodynamic pressure is defined as \cite{Kubiznak:2012wp}
\begin{equation}
P = -\frac{\Lambda}{8\pi},
\label{pressure}
\end{equation}
with the corresponding conjugate quantity identified as the thermodynamic volume,
\begin{equation}
V = \frac{4\pi r_+^3}{3},
\label{V0}
\end{equation}
where $r_+$ denotes the radius of the event horizon.

For a static, electrically charged black hole, the first law of thermodynamics in the extended phase space reads
\begin{equation}
dM = T_H\, dS + \Phi_H\, dQ + V\, dP,
\label{eq:first_law}
\end{equation}
where $T_H$ is the Hawking temperature, $S$ is the entropy, $\Phi_H$ is the electric potential at the horizon, and $Q$ is the electric charge. The associated thermodynamic quantities follow from
\begin{equation}
T_H = \left(\frac{\partial M}{\partial S}\right)_{Q,P}, 
\qquad 
\Phi_H = \left(\frac{\partial M}{\partial Q}\right)_{S,P}, 
\qquad 
V = \left(\frac{\partial M}{\partial P}\right)_{S,Q}.
\label{eq:TH_Ph}
\end{equation}

Imposing the horizon condition $f(r_+) = 0$ on the metric function in Eq.~(\ref{metricfunction}), the black hole mass can be written as
\begin{equation}
M = \frac{-3\mu Q^4 + 60 Q^2 r_+^4 - 20 \Lambda r_+^8 + 60 r_+^6}{120 r_+^5},
\label{eq:mass}
\end{equation}
where $\mu$ encodes the nonlinear electrodynamics (Euler--Heisenberg) correction.

The Hawking temperature is obtained from the surface gravity $\kappa$,
\begin{equation}
T_H = \frac{\kappa}{2\pi}, 
\qquad 
\kappa = \frac{f'(r)}{2}\Big|_{r=r_+},
\end{equation}
which yields
\begin{equation}
T_H = \frac{\mu Q^4 - 4 Q^2 r_+^4 - 4 \Lambda r_+^8 + 4 r_+^6}{16\pi r_+^7}.
\label{eq:TH}
\end{equation}
In the limit $\mu \to 0$, this expression reduces to the standard Reissner--Nordström--AdS result, as expected.

\begin{figure}[t!]
      	\centering{
      	\includegraphics[scale=0.65]{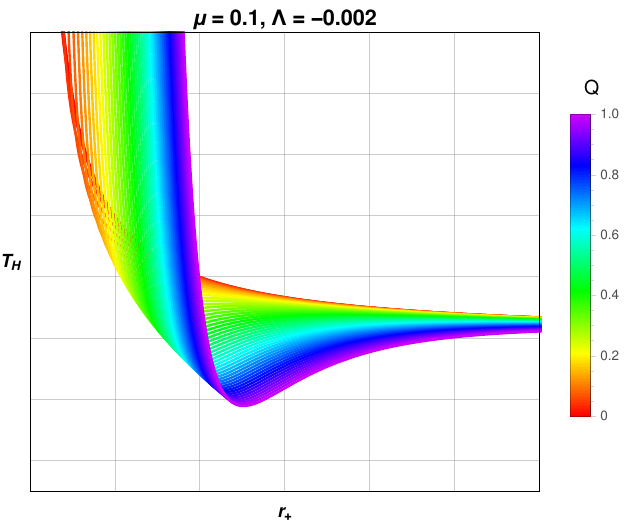} \hspace{5mm}
       \includegraphics[scale=0.65]{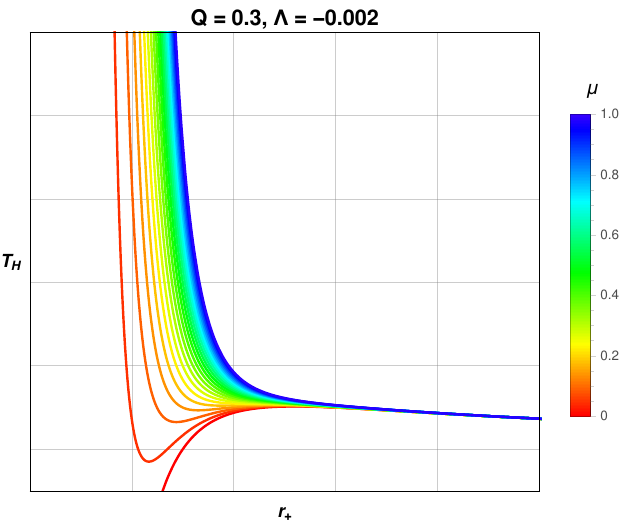}\\
       }
      	\caption{Variation of the Hawking temperature with the black hole horizon radius $r_+$.}
      	\label{figT01}
      \end{figure}

Figure~\ref{figT01} illustrates the behaviour of the Hawking temperature $T_H$ as a function of the event horizon radius $r_+$ for different values of the electric charge $Q$ and the nonlinear electrodynamics parameter $\mu$, with fixed negative cosmological constant $\Lambda<0$.

In the left panel, $\mu$ is held fixed while $Q$ is varied. For small horizon radii, the temperature diverges, reflecting the dominance of strong electromagnetic and quantum effects near the extremal regime. A local minimum in $T_H(r_+)$ appears, separating two distinct thermodynamic phases. The branch with $dT_H/dr_+<0$ corresponds to a negative heat capacity, indicating thermodynamic instability. In contrast, for sufficiently large $r_+$, the temperature increases monotonically, signalling a stable large black hole phase with positive heat capacity. Increasing the charge deepens the temperature minimum and shifts it to larger $r_+$, demonstrating that electric repulsion suppresses thermal radiation and delays the onset of stability.

In the right panel, $Q$ is fixed while $\mu$ is varied. Increasing $\mu$ enhances the nonlinear electrodynamic corrections, which effectively weaken the electric field near the horizon. This leads to an overall suppression of the Hawking temperature, particularly in the small black hole regime. Moreover, the unstable branch becomes narrower as $\mu$ increases, indicating that nonlinear electrodynamics plays a stabilizing role by smoothing high-field divergences and reducing thermal instabilities.

Overall, Fig.~\ref{figT01} reveals that the interplay between AdS curvature, electric charge, and quantum electromagnetic corrections generates a rich thermodynamic structure, characterized by multiple stability regimes and a nontrivial phase landscape.

From the first law of thermodynamics, the entropy of the black hole can be obtained from
\begin{equation}
S = \int \frac{dM}{T_H}.
\label{S0}
\end{equation}
Substituting the expression for the Hawking temperature given in Eq.~(\ref{eq:TH}) and performing the integration, one recovers the standard Bekenstein--Hawking entropy,
\begin{equation}
S = \pi r_+^2,
\end{equation}
which shows that the area law remains valid despite the presence of nonlinear electrodynamics corrections.

In the extended phase space, the black hole mass is interpreted as the enthalpy of the system. Therefore, the enthalpy is given directly by
\begin{equation}
H \equiv M.
\label{eq32}
\end{equation}
Using Eq.~(\ref{eq:mass}), the enthalpy of the black hole takes the explicit form
\begin{equation}
H = \frac{1}{8}\left(-\frac{\mu Q^4}{5 r_+^5}
+ \frac{4 Q^2}{r_+}
- \frac{4 \Lambda r_+^3}{3}
+ 4 r_+ \right).
\label{H0}
\end{equation}

The internal energy $U$ provides a measure of the total energy stored in the system excluding the work associated with the pressure--volume term. It is obtained from the standard thermodynamic relation
\begin{equation}
U = H - P V,
\end{equation}
where $P$ and $V$ are given by Eqs.~(\ref{pressure}) and (\ref{V0}), respectively. Substituting these expressions, the internal energy becomes
\begin{equation}
U = -\frac{\mu Q^4}{40 r_+^5}
+ \frac{Q^2}{2 r_+}
+ \frac{r_+}{2}.
\label{eq:U}
\end{equation}

This result shows that the internal energy receives three distinct contributions: a nonlinear electrodynamics correction proportional to $\mu$, an electrostatic term proportional to $Q^2$, and a purely gravitational term proportional to the horizon radius $r_+$. Notably, the cosmological constant does not appear explicitly in $U$, indicating that the AdS background affects the system only through the enthalpy and not through the intrinsic internal energy.

In addition to the internal energy, the Helmholtz free energy $F$ and the specific heat $C$ play central roles in determining the thermodynamic stability and phase structure of the black hole. The Helmholtz free energy,
\begin{equation}
F = U - T_H S,
\end{equation}
measures the amount of useful work that can be extracted from the system at fixed temperature, while the specific heat,
\begin{equation}
C = \left(\frac{\partial U}{\partial T_H}\right)_{Q,P},
\end{equation}
characterizes the response of the system to thermal fluctuations.

In the following sections, we analyse the behaviour of these thermodynamic potentials under small fluctuations around equilibrium. Such an analysis is crucial for understanding the stability of the Hayward--AdS black hole and identifying possible phase transitions. The combined study of internal energy, free energy, and specific heat provides a comprehensive picture of how nonlinear electrodynamics and the surrounding string fluid modify the thermodynamic landscape of the black hole system.

\section{Thermal Fluctuations and Corrected Thermodynamics}
\label{section4}

In recent years, considerable attention has been devoted to understanding quantum corrections to black hole thermodynamics. Such corrections arise naturally from thermal fluctuations and quantum effects, which become increasingly important as the black hole approaches microscopic scales. While classical black hole thermodynamics provides a robust macroscopic description, it fails to incorporate the intrinsic quantum nature of spacetime. This motivates the introduction of fluctuation-induced corrections to thermodynamic quantities, particularly entropy, in order to obtain a more accurate statistical description of small black holes.

These effects are especially relevant for asymptotically AdS black holes, where the presence of a natural thermal reservoir allows one to define a canonical ensemble and apply standard tools from equilibrium statistical mechanics. In this framework, black holes can be treated as thermodynamic systems whose microstates are governed by an underlying quantum gravitational theory.

\subsection{Corrected Entropy due to Thermal Fluctuations}

Following the seminal work of Hawking and Page \cite{Hawking:1982dh}, black holes in asymptotically AdS spacetime can be consistently described using a canonical ensemble. We therefore consider the deformed AdS black hole as a canonical system characterized by a partition function
\begin{equation}
Z = \int_0^\infty dE \, \rho(E) \, e^{-\bar{\beta}_\kappa E},
\end{equation}
where $\bar{\beta}_\kappa$ is the inverse temperature and $\rho(E)$ denotes the density of states.

Using Laplace inversion, the density of states is given by
\begin{equation}
\rho(E) = \frac{1}{2\pi i} 
\int_{\bar{\beta}_{0\kappa}-i\infty}^{\bar{\beta}_{0\kappa}+i\infty}
d\bar{\beta}_\kappa \, e^{S(\bar{\beta}_\kappa)},
\end{equation}
with the entropy
\begin{equation}
S(\bar{\beta}_\kappa) = \bar{\beta}_\kappa E + \ln Z.
\end{equation}

Expanding around the equilibrium inverse temperature $\bar{\beta}_{0\kappa}$ and noting that the first derivative vanishes at equilibrium, one obtains the corrected entropy in the presence of thermal fluctuations as
\begin{equation}
S = S_0 
- \frac{1}{2}\ln\!\left(S_0 T_H^2\right)
+ \frac{f(m,n)}{S_0}
+ \cdots ,
\end{equation}
where $S_0$ is the classical entropy and higher-order terms represent further quantum corrections \cite{More:2004hv,Gogoi:2024ypn,Gursel:2025wan}.

A more general parametrization is given by \cite{Pourhassan:2016zzc}
\begin{equation}
S = S_0 
- \lambda_1 \ln\!\left(S_0 T_H^2\right)
+ \frac{\lambda_2}{S_0}
+ \cdots ,
\end{equation}
where $\lambda_1$ and $\lambda_2$ control the strength of the first- and second-order corrections, respectively. The first term represents universal logarithmic quantum corrections, while the second term corresponds to geometric corrections suppressed by the entropy.

Including both corrections, the explicit form of the corrected entropy for the present black hole reads
\begin{equation}
S_c = \pi r_+^2 
- \lambda_1 
\ln\!\left(
\frac{\left(\mu Q^4 - 4Q^2 r_+^4 - 4\Lambda r_+^8 + 4r_+^6\right)^2}
{r_+^{12}}
\right)
+ \frac{\lambda_2}{\pi r_+^2}
+ \lambda_1 \ln(256\pi).
\label{sc}
\end{equation}
This expression reduces to the standard Bekenstein--Hawking entropy in the limit 
$\lambda_1,\lambda_2 \to 0$.

\begin{figure}[t!]
      	\centering{
      	\includegraphics[scale=0.65]{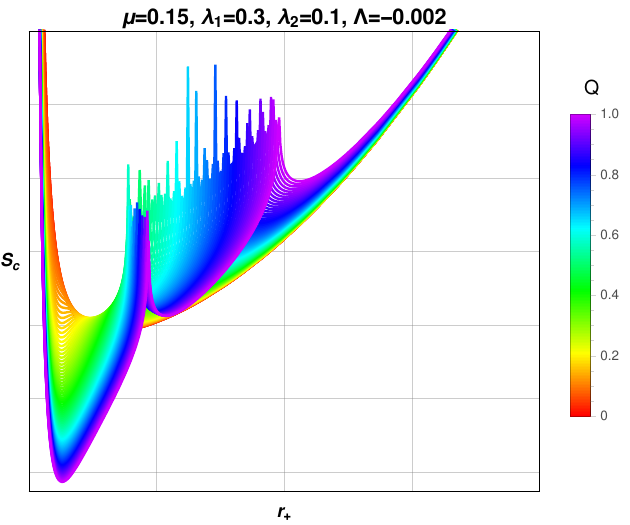} \hspace{5mm}
       \includegraphics[scale=0.65]{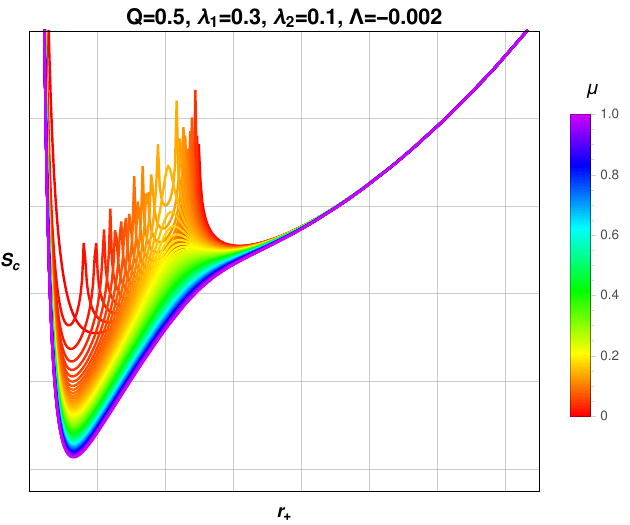}\\
       \includegraphics[scale=0.65]{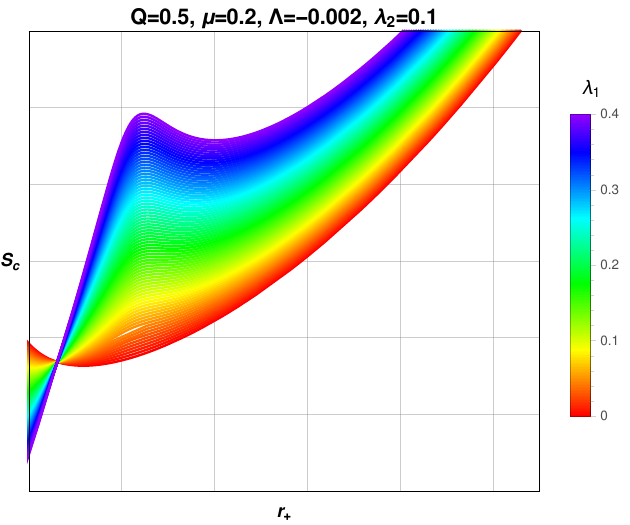}\hspace{5mm}
       \includegraphics[scale=0.65]{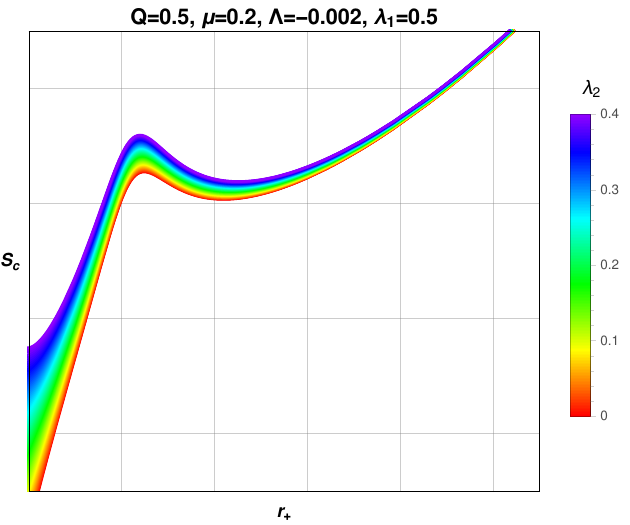}
       }
      	\caption{Variation of the black hole entropy with the horizon radius $r_+$.}
      	\label{figSc01}
      \end{figure}

Figure~\ref{figSc01} illustrates the behaviour of the corrected entropy $S_c$ as a function of the horizon radius $r_+$ for different values of the charge $Q$, nonlinear parameter $\mu$, and correction parameters $\lambda_1$ and $\lambda_2$.

In the first panel, the entropy exhibits a pronounced minimum at small horizon radii, followed by a monotonic increase for larger $r_+$. This indicates the presence of a thermodynamically unstable regime for small black holes, where quantum fluctuations dominate. As $Q$ increases, oscillatory structures appear near the minimum, signalling competing quantum and electromagnetic effects. These oscillations are typical signatures of phase transitions or metastable states in quantum gravitational systems.

In the second panel, increasing $\mu$ systematically raises the entropy profile. Since $\mu$ weakens the effective electric field through nonlinear electrodynamics, this leads to enhanced microscopic degeneracy and hence larger entropy. The disappearance of sharp oscillations at large $\mu$ reflects improved thermodynamic stability.

In the third panel, the influence of $\lambda_1$ is shown to be dominant. Larger $\lambda_1$ amplifies logarithmic corrections, producing strong deviations from the area law at small $r_+$. This confirms that $\lambda_1$ governs the quantum sector of the theory. In contrast, variations of $\lambda_2$ generate only mild shifts in $S_c$, demonstrating that second-order geometric corrections are subleading.

Notably, for very small horizon radii, $S_c$ can become negative in certain parameter regimes. Negative entropy signals a breakdown of the thermodynamic description and marks the onset of a nonphysical or unstable phase. Beyond a critical radius, the entropy becomes positive and grows monotonically, indicating restoration of classical thermodynamic behaviour.

Overall, Fig.~\ref{figSc01} reveals that thermal fluctuations generate a rich entropy landscape, characterized by unstable quantum-dominated phases for microscopic black holes and stable classical phases for macroscopic ones. The transition between these regimes is controlled primarily by $\lambda_1$ and $\mu$, highlighting the central role of quantum and nonlinear electromagnetic effects in black hole thermodynamics.

\subsection{Corrected Potentials due to Thermal Fluctuations}
\label{subsec:corrected_potentials}

In this subsection, we investigate the influence of thermal fluctuations on the thermodynamic potentials of the Hayward--AdS black hole. Using the corrected entropy derived in the previous section, the corresponding corrected enthalpy $H_c$ can be obtained from the first law of thermodynamics in the extended phase space,
\begin{equation}
H_c = \int T_H\, dS_c, 
\label{Enthalpy}
\end{equation}
where $V_c$ denotes the corrected thermodynamic volume. Since the correction parameters $\lambda_1$ and $\lambda_2$ are taken to be constants, the second term does not contribute explicitly and can be neglected. Substituting the expressions for $T_H$ and $S_c$ from Eqs.~(\ref{eq:TH}) and (\ref{sc}), we obtain
\begin{equation}
\begin{aligned}
H_c =& -\frac{\mu Q^4}{40 r_+^5}
+ \frac{Q^2}{2 r_+}
- \frac{\Lambda r_+^3}{6}
+ \frac{r_+}{2} \\
&+ \frac{30\pi \lambda_1 r_+^2 \left(-9\mu Q^4 + 28 Q^2 r_+^4 + 84\Lambda r_+^8\right)
+ 7\lambda_2 \left(5\mu Q^4 - 36 Q^2 r_+^4 - 180\Lambda r_+^8 + 60 r_+^6\right)}
{2520\pi^2 r_+^9}.
\end{aligned}
\end{equation}

The first four terms correspond to the classical enthalpy, while the remaining terms encode quantum corrections due to thermal fluctuations. These corrections become relevant primarily in the small-horizon regime, where quantum effects dominate.

\begin{figure}[t!]
      	\centering{
      	\includegraphics[scale=0.65]{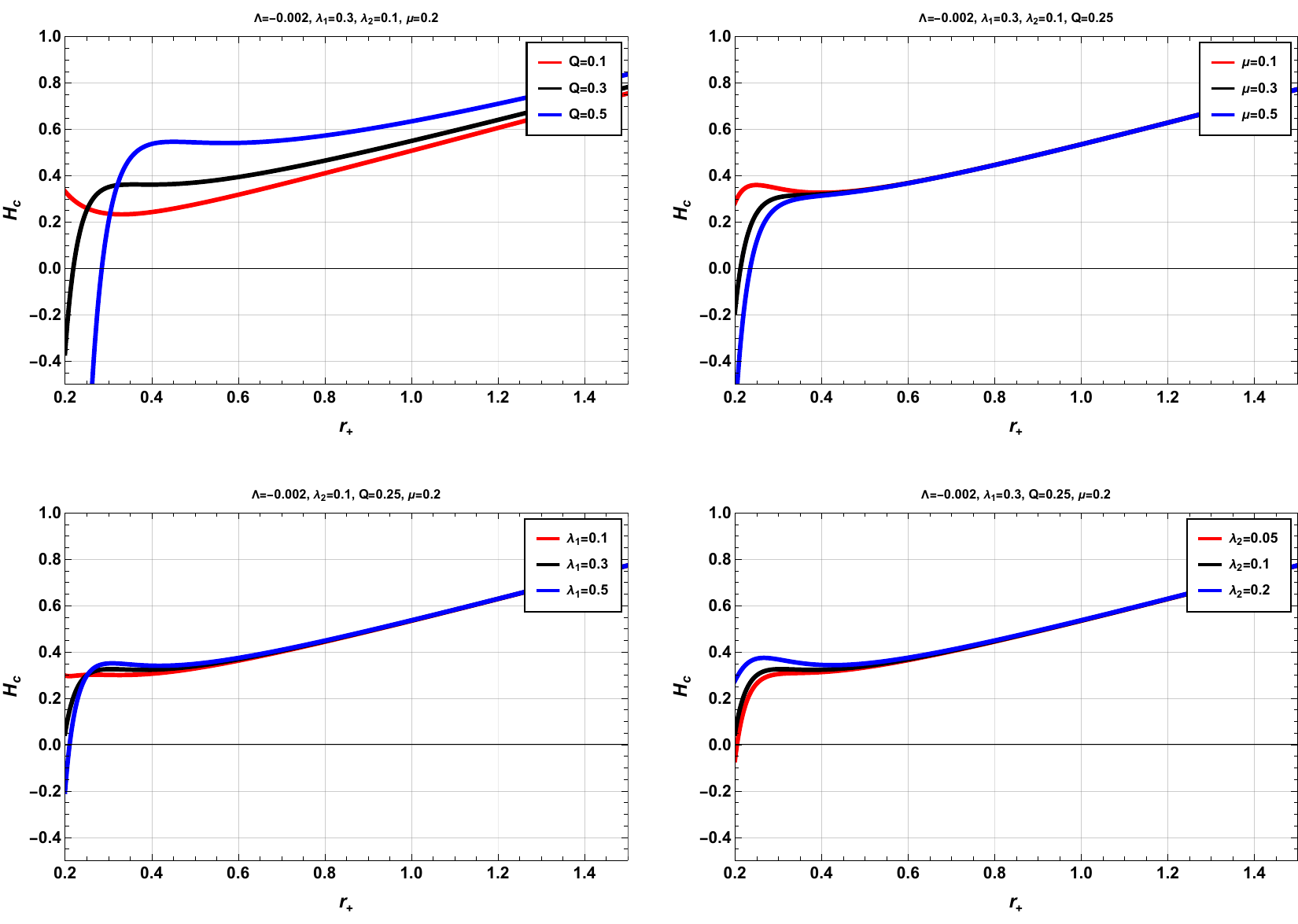} }
      	\caption{Variation of the black hole's corrected enthalpy $H_c$ with the horizon radius $r_+$.}
      	\label{hc01}
      \end{figure}

Figure~\ref{hc01} shows the variation of the corrected enthalpy $H_c$ as a function of the horizon radius $r_+$ for different values of the system parameters.

In the first panel, increasing the charge $Q$ leads to a systematic increase in $H_c$ for a fixed horizon radius. This reflects the additional electrostatic energy stored in the black hole. A shallow minimum appears at small $r_+$, indicating a critical radius separating unstable microscopic black holes from stable macroscopic ones. For large $r_+$, the enthalpy grows monotonically, confirming the thermodynamic stability of large AdS black holes.

In the second panel, variations in the nonlinear parameter $\mu$ modify the small-radius behaviour significantly. Larger values of $\mu$ suppress the enthalpy in the near-extremal regime, demonstrating that nonlinear electrodynamics weakens the effective electric field and reduces the total energy content. However, at large $r_+$ all curves converge, showing that quantum corrections become negligible for macroscopic black holes.

In the third panel, increasing the first-order quantum correction parameter $\lambda_1$ slightly lowers the enthalpy curve. This confirms that logarithmic corrections effectively reduce the accessible energy of the system, acting as a quantum backreaction on the classical geometry.

In the fourth panel, variations in $\lambda_2$ produce only marginal changes in $H_c$. The near overlap of the curves indicates that second-order corrections are subleading and have little impact on the global thermodynamic behaviour.

Overall, Fig.~\ref{hc01} demonstrates that the corrected enthalpy exhibits a stable monotonic growth for large black holes, while quantum and nonlinear effects dominate the thermodynamics of small black holes. The presence of a minimum in $H_c$ signals the existence of a critical scale below which classical thermodynamics breaks down.

Finally, the corrected thermodynamic volume follows from
\begin{equation}
V_c = \left(\frac{\partial H_c}{\partial P}\right)_{S_c},
\label{vc}
\end{equation}
which yields
\begin{equation}
V_c = \frac{4\pi r_+^3}{3}
- 8\lambda_1 r_+
+ \frac{4\lambda_2}{\pi r_+}.
\label{V_c}
\end{equation}
This expression shows that quantum fluctuations modify the effective volume of the black hole, introducing both linear and inverse-radius corrections. These deviations become significant only at small scales, while the classical volume is recovered for large $r_+$.

We now examine the influence of thermal fluctuations on the Helmholtz free energy of the black hole. The Helmholtz free energy plays a crucial role in characterizing thermodynamic stability, since it measures the amount of useful work that can be extracted from the system at fixed temperature and volume.

The classical (uncorrected) Helmholtz free energy is defined by
\begin{equation}
F = U - T_H S,
\end{equation}
which yields
\begin{equation}
F = \frac{1}{16} 
\left(
-\frac{7\mu Q^4}{5 r_+^5}
+ \frac{12 Q^2}{r_+}
+ \frac{4\Lambda r_+^3}{3}
+ 4 r_+
\right).
\label{F}
\end{equation}

When thermal fluctuations are taken into account, the corrected Helmholtz free energy can be obtained from the standard thermodynamic relation
\begin{equation}
F_c = -\int S_c \, dT_H - \int P \, dV_c,
\label{Fc}
\end{equation}
where $S_c$ and $V_c$ denote the corrected entropy and volume, respectively.

Substituting the expressions for $S_c$, $T_H$, and $V_c$, the corrected Helmholtz free energy becomes

\begin{equation}
\begin{aligned}
         F_c =&\frac{63 \pi ^2 r_+^4 f_{\text{c0}}+15 \pi  \lambda _1 r_+^2 \left(4 Q^2 \left(28 r_+^4-9 \mu  Q^2\right)-21 \log (256 \pi ) f_{\text{c1}}\right)+315 \pi  \lambda _1 r_+^2 f_{\text{c1}} \log \left(\frac{f_{\text{c1}}^2}{r_+^{12}}\right)+7 \lambda _2 f_{\text{c2}}}{5040 \pi ^2 r_+^9},
\end{aligned}
\end{equation}
where
$$f_{\text{c0}}=20 r_+^4 \left(3 Q^2+\Lambda  r_+^4+r_+^2\right)-7 \mu  Q^4,$$
$$f_{\text{c1}}=\mu  Q^4-4 Q^2 r_+^4-4 \Lambda  r_+^8+4 r_+^6,$$
$$f_{\text{c2}}=-35 \mu  Q^4+108 Q^2 r_+^4+180 \Lambda  r_+^8-60 r_+^6.$$

This expression reduces to the classical Helmholtz free energy in the limit 
$\lambda_1,\lambda_2 \rightarrow 0$, confirming the consistency of the formulation.

\begin{figure}[t!]
      	\centering{
      	\includegraphics[scale=0.65]{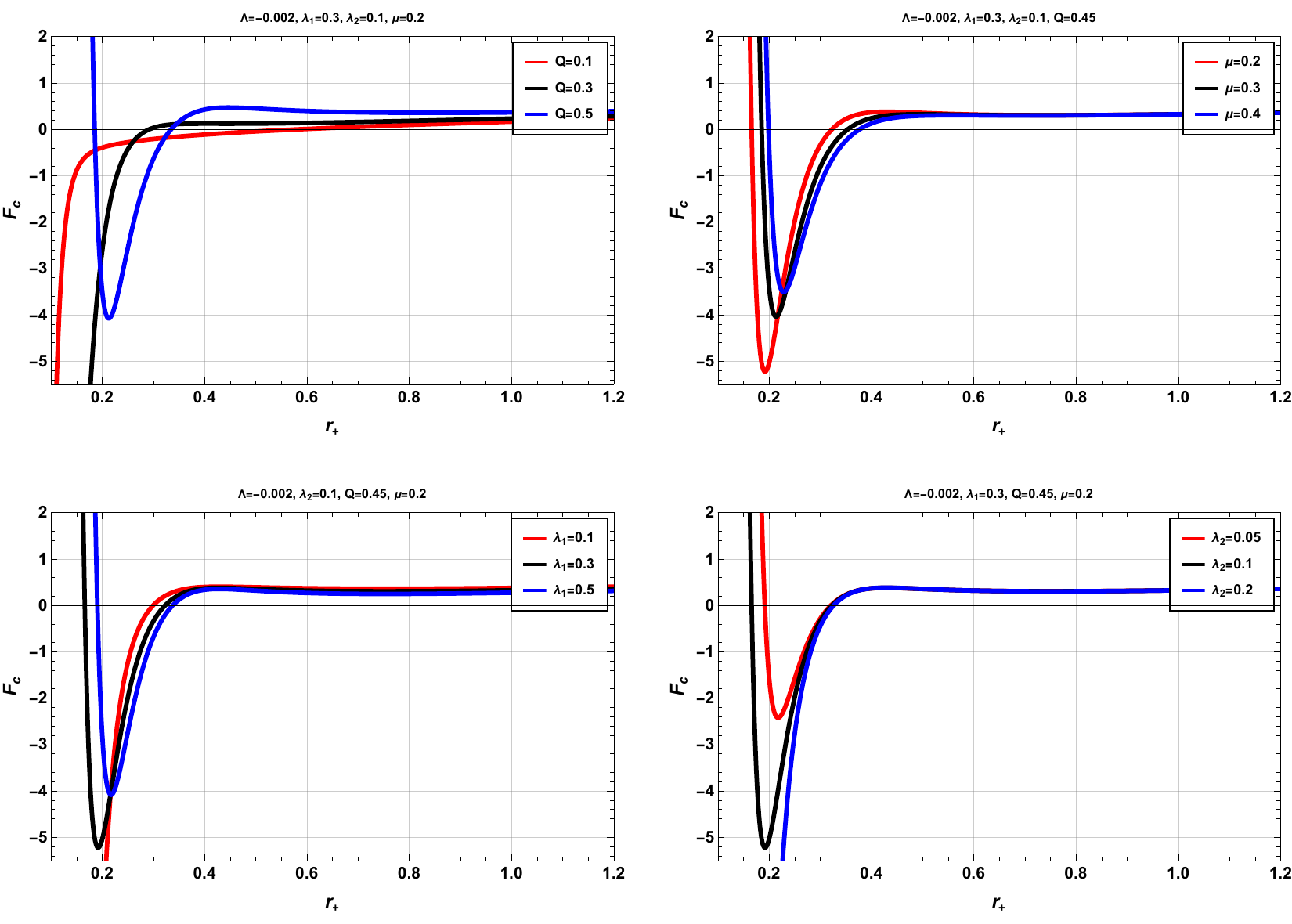}  }
      	\caption{Variation of the black hole's Helmholtz free energy $F_c$ with the horizon radius $r_+$.}
      	\label{fc}
      \end{figure}
Figure~\ref{fc} depicts the variation of the corrected Helmholtz free energy $F_c$ as a function of the horizon radius $r_+$ for different values of the charge $Q$, nonlinear electrodynamics parameter $\mu$, and correction parameters $\lambda_1$ and $\lambda_2$.

In the top-left panel, increasing the electric charge $Q$ deepens the minimum of $F_c$ at small horizon radii. This indicates that charged black holes possess a larger negative free energy in the microscopic regime, making them energetically favoured but thermodynamically unstable. The presence of a pronounced minimum suggests the existence of a critical radius below which quantum effects dominate, and the system undergoes a phase transition. At larger $r_+$, all curves approach positive values, signalling the stabilization of macroscopic black holes.

In the top-right panel, increasing the nonlinear parameter $\mu$ suppresses the depth of the free energy well. This behaviour reflects the weakening of the effective electric field due to nonlinear electrodynamics, which reduces the magnitude of negative free energy and improves the thermodynamic stability of small black holes. The convergence of the curves at large $r_+$ confirms that nonlinear corrections become negligible for macroscopic systems.

In the bottom-left panel, the first-order quantum correction parameter $\lambda_1$ significantly affects the small-radius behaviour. Larger $\lambda_1$ values enhance the depth of the free energy minimum, indicating stronger quantum backreaction and greater departure from classical thermodynamics. This demonstrates that logarithmic corrections dominate the microscopic phase structure of the black hole.

In the bottom-right panel, variations in the second-order correction parameter $\lambda_2$ produce only minor changes in $F_c$. The near-overlapping curves show that geometric corrections are subleading compared to logarithmic quantum corrections and have little impact on the global thermodynamic behaviour.

Overall, Fig.~\ref{fc} reveals that the Helmholtz free energy exhibits a characteristic quantum-dominated phase at small horizon radii and a classical stable phase at large radii. The transition between these regimes is governed primarily by the electric charge and the logarithmic quantum correction parameter $\lambda_1$, while nonlinear electrodynamics acts to moderate quantum instabilities.

Quantum and thermal fluctuations modify not only the entropy and enthalpy of the system, but also other thermodynamic potentials such as the internal energy and the Gibbs free energy. These quantities play a central role in determining the stability and phase structure of the black hole.

The corrected internal energy is obtained from the standard thermodynamic relation
\begin{equation}
U_c = H_c - P V_c ,
\label{uc}
\end{equation}
where $H_c$ and $V_c$ denote the corrected enthalpy and volume, respectively. Substituting their explicit forms, the corrected internal energy becomes
\begin{equation}
\begin{aligned}
U_c = 
\frac{
\mu Q^4 \left(35\lambda_2 - 9\pi r_+^2 (30\lambda_1 + 7\pi r_+^2)\right)
+ 84 r_+^4 
\left(
-3\lambda_2 Q^2 
+ 10\pi \lambda_1 Q^2 r_+^2 
+ 15\pi^2 r_+^4 (Q^2 + r_+^2) 
+ 5\lambda_2 r_+^2
\right)
}
{2520 \pi^2 r_+^9}.
\end{aligned}
\end{equation}

The Gibbs free energy, which determines the maximum mechanical work obtainable from the system at fixed temperature and pressure, is defined as
\begin{equation}
G = M - T_H S.
\end{equation}
The uncorrected expression takes the form
\begin{equation}
G = -\frac{7\mu Q^4}{80 r_+^5}
+ \frac{3Q^2}{4 r_+}
+ \frac{r_+}{12}\left(\Lambda r_+^2 + 3\right).
\end{equation}

Including thermal fluctuations through the corrected entropy $S_c$, the corrected Gibbs free energy becomes
\begin{equation}
\begin{aligned}
G_c =& 
-\frac{f_{\mathrm{c1}}}{16\pi r_+^7}
\left[
-\lambda_1 
\ln\!\left(\frac{f_{\mathrm{c1}}^2}{r_+^{12}}\right)
+ \lambda_1 \ln(256\pi)
+ \frac{\lambda_2}{\pi r_+^2}
+ \pi r_+^2
\right] \\
& -\frac{\mu Q^4}{40 r_+^5}
+ \frac{Q^2}{2 r_+}
- \frac{\Lambda r_+^3}{6}
+ \frac{r_+}{2},
\end{aligned}
\end{equation}
where
\begin{equation}
f_{\mathrm{c1}} = \mu Q^4 - 4Q^2 r_+^4 - 4\Lambda r_+^8 + 4 r_+^6 .
\end{equation}

This expression reduces to the classical Gibbs free energy in the limit 
$\lambda_1, \lambda_2 \rightarrow 0$, confirming the internal consistency of the corrected formalism.

\begin{figure}[t!]
      	\centering{
      	\includegraphics[scale=0.65]{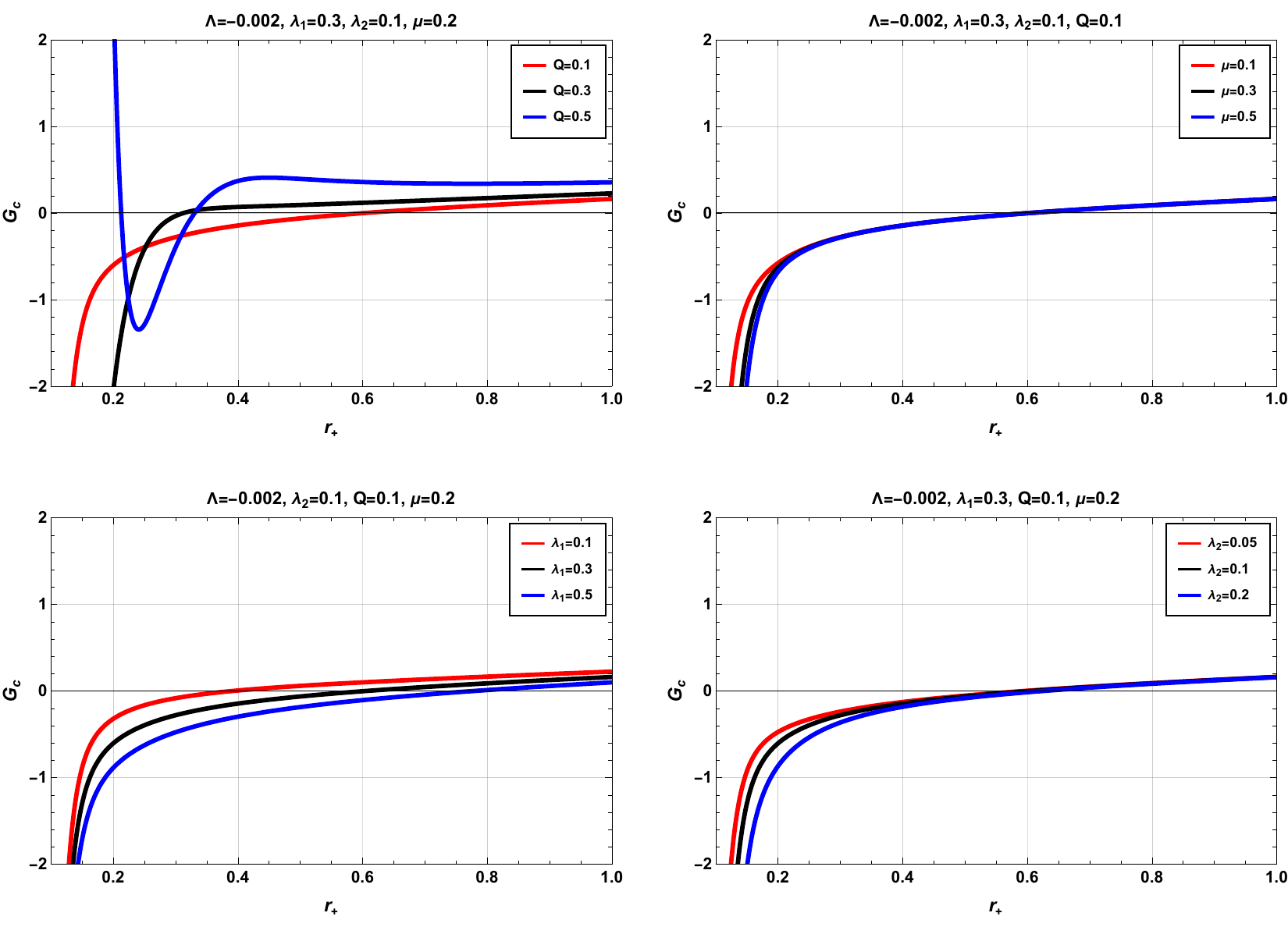} }
      	\caption{Variation of the black hole $G_c$ with the horizon radius $r_+$using $\Lambda=-0.002$.}
      	\label{figGc01}
      \end{figure}

Figure~\ref{figGc01} displays the variation of the corrected Gibbs free energy $G_c$ as a function of the horizon radius $r_+$ for different values of the charge $Q$, nonlinear electrodynamics parameter $\mu$, and correction parameters $\lambda_1$ and $\lambda_2$, with the cosmological constant fixed at $\Lambda=-0.002$.

In the top-left panel, the dependence of $G_c$ on the electric charge $Q$ is illustrated. For small horizon radii, $G_c$ is negative and exhibits a pronounced minimum, indicating a highly unstable quantum-dominated phase. As the charge increases, the depth of this minimum becomes larger, showing that electric charge enhances the magnitude of negative Gibbs free energy and thus amplifies thermodynamic instability in the microscopic regime. For larger values of $r_+$, all curves increase monotonically and gradually approach positive values, signifying the restoration of thermodynamic stability for macroscopic black holes.

In the top-right panel, the effect of the nonlinear parameter $\mu$ is shown. Increasing $\mu$ shifts the curves upward and reduces the depth of the negative region at small $r_+$. This behaviour reflects the weakening of the effective electric field due to nonlinear electrodynamics, which suppresses extreme quantum effects and stabilizes small black holes. At large $r_+$, the curves corresponding to different $\mu$ nearly coincide, indicating that nonlinear corrections become negligible in the classical regime.

In the bottom-left panel, the influence of the first-order quantum correction parameter $\lambda_1$ is presented. Larger values of $\lambda_1$ significantly lower the Gibbs free energy, especially in the small-radius region. This demonstrates that logarithmic quantum corrections dominate the thermodynamic behaviour of microscopic black holes and strongly modify the amount of mechanical work extractable from the system.

In the bottom-right panel, the variation of $G_c$ with respect to the second-order correction parameter $\lambda_2$ is shown. The curves almost overlap, with only minor deviations near small $r_+$. This indicates that second-order geometric corrections have a comparatively weak effect on the Gibbs free energy and play a subleading role in the thermodynamic structure.

Overall, Fig.~\ref{figGc01} reveals a clear separation between two thermodynamic regimes: a quantum-dominated unstable phase for small horizon radii characterized by negative Gibbs free energy, and a classical stable phase for large horizon radii where $G_c$ becomes positive and varies smoothly. The transition between these regimes is governed primarily by the electric charge and the logarithmic quantum correction parameter $\lambda_1$, while nonlinear electrodynamics acts to moderate quantum instabilities.

\section{Phase Transitions and Stability}
\label{section5}

The thermodynamic stability of a black hole is primarily determined by its specific heat. The specific heat measures the response of the system to thermal perturbations and is defined as
\begin{equation}
C = T \left(\frac{\partial S}{\partial T}\right),
\end{equation}
so that its sign directly encodes stability. A positive specific heat indicates that the system absorbs heat and warms up in a controlled manner, corresponding to a locally stable thermodynamic configuration. In contrast, a negative specific heat implies runaway behaviour, where the system heats up as it loses energy, signalling thermodynamic instability.

For the present black hole, the uncorrected specific heat is given by
\begin{equation}
C_0 = 
\frac{2 \pi r_+^2 
\left(
-\mu Q^4 + 4Q^2 r_+^4 + 4\Lambda r_+^8 - 4r_+^6
\right)}
{7\mu Q^4 - 12Q^2 r_+^4 + 4\left(\Lambda r_+^8 + r_+^6\right)}.
\end{equation}
This expression reproduces the standard behaviour of charged AdS black holes in the absence of quantum corrections, where stability is governed by a competition between gravitational attraction, electromagnetic repulsion and the AdS confining potential.

When thermal fluctuations are incorporated, the corrected specific heat takes the form
\begin{equation}
C_c = 
\frac{
2\mu Q^4\left(\lambda_2 - \pi r_+^2(6\lambda_1 + \pi r_+^2)\right)
+ 8\pi r_+^6
\left[
2\lambda_1(Q^2 - \Lambda r_+^4)
+ \pi r_+^2(Q^2 + \Lambda r_+^4 - r_+^2)
\right]
- 8\lambda_2 r_+^4(Q^2 + \Lambda r_+^4 - r_+^2)
}
{
7\pi\mu Q^4 r_+^2
+ 4\pi r_+^6(-3Q^2 + \Lambda r_+^4 + r_+^2)
}.
\end{equation}
The presence of the correction parameters $\lambda_1$ and $\lambda_2$ introduces new microscopic contributions to the heat capacity, which become dominant in the small-horizon regime.

\begin{figure}[t!]
      	\centering{
      	\includegraphics[scale=0.65]{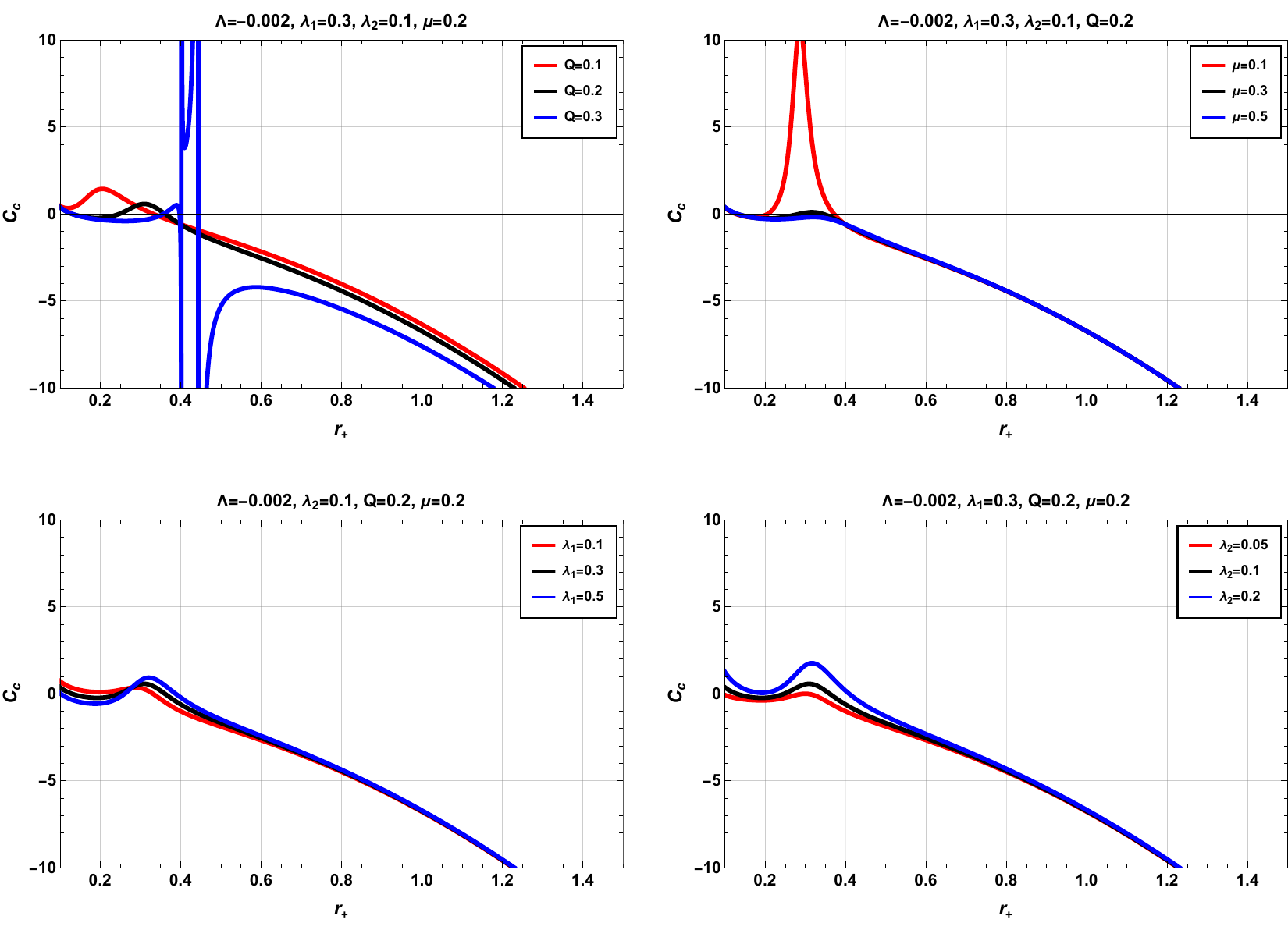}
       }
      	\caption{Variation of the black hole $C_c$ with the horizon radius $r_+$ using $\Lambda=-0.002$. }
      	\label{figCc01}
      \end{figure}

The behaviour of $C_c$ as a function of the horizon radius is displayed in Fig.~\ref{figCc01}.

\subsection*{Phase Structure and Stability}

Figure~\ref{figCc01} shows that the corrected specific heat exhibits multiple sign changes and divergences. These features signal the existence of several thermodynamic phases and genuine phase transitions. In particular:

\begin{itemize}
\item Regions where $C_c>0$ correspond to locally stable black hole phases.
\item Regions where $C_c<0$ correspond to unstable phases.
\item Divergences of $C_c$ correspond to second-order phase transitions, where thermal fluctuations diverge and the system becomes critically unstable.
\end{itemize}

{ Unlike the classical Reissner--Nordström--AdS case, where large black holes are typically stable in the canonical ensemble, the present system exhibits a much richer structure. The divergences in the corrected specific heat $C_c$ indicate genuine continuous (second-order) phase transitions. Each divergence occurs at a dynamically determined critical horizon radius, $r_c$, which serves as the boundary separating distinct thermodynamic phases. As the horizon radius $r_+$ grows and crosses these critical points, the system undergoes a phase transition from an initial quantum-stabilized state to a final thermodynamically unstable state (or vice versa, depending on the parameter regime).

To complete this discussion and provide clear definitions, we classify black holes as ``small'' (microscopic) or ``large'' (macroscopic) with respect to these critical radii:
\begin{itemize}
    \item \textbf{Small (Microscopic) Black Holes:} These correspond to black holes with a horizon radius below the first critical radius ($r_+ < r_c$). In this regime, logarithmic quantum corrections dominate, generating narrow windows of positive specific heat ($C_c > 0$). This initial state reflects a \emph{quantum-stabilized microscopic phase}.
    \item \textbf{Large (Macroscopic) Black Holes:} These correspond to black holes with a horizon radius extending beyond the outermost critical radius ($r_+ > r_c$). In this final state, the specific heat becomes universally negative ($C_c < 0$). This implies that, despite the confining AdS background, classical macroscopic instability takes over, and the quantum corrections become negligible.
\end{itemize}

This behaviour is consistent with the analysis of the corrected Helmholtz and Gibbs free energies in Section \ref{section4}, where large black holes were found to possess negative free energy slopes and no stable equilibrium branch.}

The role of individual parameters is physically transparent:

\begin{itemize}
\item The electric charge $Q$ enhances the magnitude of divergences in $C_c$, producing multiple critical points. This reflects the destabilizing role of electromagnetic repulsion, which amplifies thermal fluctuations.
\item The nonlinear electrodynamics parameter $\mu$ suppresses the positive specific heat regions, indicating that nonlinear field effects reduce thermal stability by weakening the effective confining force.
\item The logarithmic correction parameter $\lambda_1$ plays a dominant role in generating stable phases, showing that first-order quantum corrections are primarily responsible for stabilizing microscopic black holes.
\item The second-order correction parameter $\lambda_2$ introduces only minor quantitative shifts, confirming that higher-order geometric corrections are subleading.
\end{itemize}

Thus, the phase structure inferred from $C_c$ is fully consistent with the global thermodynamic picture obtained from entropy, enthalpy and free energies: quantum effects dominate small black holes, while classical instabilities dominate large ones.

\section{Conclusion}

In this work, we have investigated the thermodynamic behaviour of an AdS black hole arising from nonlinear electrodynamics-corrected gravity, incorporating quantum corrections induced by thermal fluctuations. Starting from the Einstein--Euler--Heisenberg framework, we derived the modified black hole solution and systematically analysed its thermodynamic properties in the extended phase space, where the cosmological constant is interpreted as thermodynamic pressure.

We first obtained the fundamental thermodynamic quantities, including the corrected entropy, enthalpy, internal energy, Helmholtz free energy and Gibbs free energy. The entropy was shown to receive logarithmic and inverse-area corrections, governed by the parameters $\lambda_1$ and $\lambda_2$, which encode first- and second-order quantum effects. While the area law remains valid in the classical limit, quantum corrections become significant in the small-horizon regime, leading to strong deviations from standard black hole thermodynamics. The corrected enthalpy and internal energy revealed that nonlinear electrodynamics and thermal fluctuations modify the energy content of the black hole primarily at microscopic scales. In particular, the logarithmic correction parameter $\lambda_1$ plays a dominant role, whereas the second-order parameter $\lambda_2$ introduces only subleading modifications. This hierarchy was consistently reflected across all thermodynamic potentials.
The analysis of the Helmholtz and Gibbs free energies demonstrated that small black holes are quantum dominated and exhibit deep free energy minima, while large black holes approach a smooth classical regime. The free energy structure indicates that quantum corrections reshape the thermodynamic landscape by introducing new metastable configurations and modifying the amount of mechanical work extractable from the system.

A central result of this work is the behaviour of the corrected specific heat. We found that thermal fluctuations generate multiple sign changes and divergences in the heat capacity, signalling the presence of genuine second-order phase transitions. Unlike classical AdS black holes, where large black holes are typically stable, the corrected system exhibits narrow stable windows at small horizon radii and universal instability at large scales. This shows that quantum effects qualitatively alter the stability structure, producing a quantum-stabilized microscopic phase followed by a classically unstable macroscopic phase. The role of individual physical parameters is transparent. Electric charge amplifies thermal instabilities and generates multiple critical points. The nonlinear electrodynamics parameter $\mu$ suppresses stability by weakening the effective confining force. The logarithmic correction parameter $\lambda_1$ governs the emergence of stable quantum phases, while $\lambda_2$ remains subdominant throughout.

Our results demonstrate that thermal fluctuations do not merely introduce small perturbative corrections, but fundamentally reorganizes the phase structure of black holes in nonlinear electrodynamics. The thermodynamic behaviour is governed by a clear quantum--classical transition: microscopic black holes are controlled by quantum fluctuations and exhibit rich phase behaviour, whereas macroscopic black holes are dominated by classical instabilities.

These findings provide further evidence that black hole thermodynamics, when corrected by quantum effects, encodes deep information about the microstructure of spacetime. In particular, the emergence of logarithmic corrections and quantum-induced phase transitions suggests that black holes act as sensitive probes of underlying quantum gravity effects. Our study therefore strengthens the view that thermal fluctuations play a crucial role in bridging classical gravity and quantum spacetime physics.

\section*{Acknowledgment}
 DJG acknowledges the contribution of the COST Action CA21136  -- ``Addressing observational tensions in cosmology with systematics and fundamental physics (CosmoVerse)". 

\section*{Declaration of competing interest}
The authors declare that they have no known competing financial interests or personal relationships that could have appeared to influence the work reported in this manuscript.

\section*{Data Availability Statement}
There are no new data associated with this article.

\section*{Funding Information}
Not applicable.

\bibliographystyle{apsrev}
\bibliography{refs}
\end{document}